 \documentclass[final,3p,times,twocolumn]{elsarticle}

\usepackage{amssymb}

\journal{Physics Letters B}

\usepackage{amsmath}
\usepackage{slashed}
\usepackage{amssymb}
\usepackage{braket}
\usepackage{graphicx}
\usepackage{amsmath}
\usepackage{braket}
\usepackage{epigraph}
\usepackage{tensor}
\usepackage{enumerate}
\usepackage{subfig}
\usepackage{verbatim}
\usepackage{pstricks}
\usepackage{epstopdf}
\usepackage{tikz}
\usepackage{pgfplots}
\pgfplotsset{/pgf/number format/use comma,compat=newest}
\usepackage{color}

\usepackage{feynmp}
\makeatletter
\def\endfmffile{%
  \fmfcmd{\p@rcent\space the end.^^J%
          end.^^J%
          endinput;}%
  \if@fmfio
    \immediate\closeout\@outfmf
  \fi
  \ifnum\pdfshellescape=\@ne
    \immediate\write18{mpost \thefmffile}%
  \fi}
\makeatother

\begin{document}

\begin{frontmatter}


\author{J. I. McDonald}
\ead{pymcdonald@swansea.ac.uk}
\author{G. M. Shore}
\ead{g.m.shore@swansea.ac.uk}
\address{Department of Physics, College of Science, \\
Swansea University, Singleton Park, Swansea, SA2 8PP.}

%


\title{Leptogenesis and gravity:  baryon asymmetry without decays}


\author{}

\address{}

\begin{abstract}
A popular class of theories attributes the matter-antimatter asymmetry of the Universe to CP-violating decays of super-heavy BSM particles in the Early Universe. Recently, we discovered a new source of leptogenesis in these models, namely that the same Yukawa phases which provide the CP violation for decays, combined with curved-spacetime loop effects, lead to an entirely new gravitational mechanism for generating an asymmetry, driven by the expansion of the Universe and independent of the departure of the heavy particles from equilibrium. In this Letter, we build on previous work by analysing the full Boltzmann equation, exploring the full parameter space of the theory and studying the time-evolution of the asymmetry. Remarkably, we find regions of parameter space where decays play no part at all, and where the baryon asymmetry of the Universe is determined solely by gravitational effects.
\end{abstract}

\end{frontmatter}



\section{Introduction}
In a series of recent papers \cite{RIGLletter,McDonaldShore:Dec2015} we described a new phenomenon whereby gravity drives the Universe towards a matter-antimatter asymmetry. Our main realisation was that matter and antimatter propagate differently in the presence of gravity when CP symmetry is violated. Specifically, we proved \cite{RIGLletter,McDonaldShore:Dec2015} that in translation invariant environments, CPT symmetry necessarily forces matter and antimatter to propagate identically. Conversely, when this symmetry is broken by the background geometry, \textit{e.g.,} an expanding Universe, and when there is a source of CP violation, matter/antimatter propagators become distinct. This causes a spectral splitting for matter/antimatter and an energy cost difference which drives the system towards an asymmetric state, facilitated by particle number-violating reactions.

\begin{figure}
\centering
\includegraphics[scale=0.42]{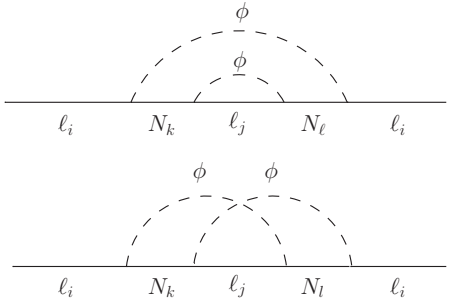}
\caption{Loop diagrams which give distinct matter/antimatter propagators and which generate the operator (\ref{Operator}).}
\label{2loop}
\end{figure}

As in our previous papers, we shall illustrate this effect within the context of leptogenesis \cite{FukYan}, though as will become apparent, it applies equally well in any theory with a source of CP violation and B or L violation. In this case, the Lagrangian -- minimally coupled to gravity -- is given by
\begin{align}
\mathcal{L} =\sqrt{-g}\left[ \overline{N} \slashed{D} N+\overline{N}
\, M \,  N+ h_{i j} \bar{\ell}_i \phi  N_j    + \mbox{h.c.}\right]   ,
\label{FYmodel}
\end{align} 
where $\ell_i$ are the left-handed lepton doublets, $\phi$ is the charge-conjugate Higgs doublet, and $N_i$ are sterile neutrinos, written here in the Majorana basis\footnote{In previous papers \cite{RIGLletter,McDonaldShore:Dec2015}, as in \cite{FukYan}, we used $N$ to label the basis of RH neutrinos, which are now more usually denoted $(\nu)_R$. } so that $N^c = N$. As described above, at two-loops (figure \ref{2loop}) in a time-dependent gravitational background, lepton and antilepton self-energies are distinct $\Sigma_{\ell}(x,x') \neq \Sigma_{\bar{\ell}}(x,x')$.

Minimal coupling ensures that at tree-level, the strong equivalence principle holds and leptons are insensitive to curvature, but when loop effects are taken into account, two things happen. Firstly, the propagators become sensitive to CP violation contained in the Yukawa couplings, a symmetry which obviously must be broken for distinct propagation. Moreover, as described in \cite{Drummond:1979pp,Shore:2004sh} the screening cloud surrounding the propagating leptons causes them to acquire an effective ``size" and experience gravitational tidal forces, violating the strong equivalence principle and causing the leptons to couple directly to curvature. 

When the sterile neutrinos are integrated out from the diagrams in figure \ref{2loop}, the resulting effective action contains the following CP- and strong equivalence principle-violating operator for each lepton generation:
\begin{equation}\label{Operator}
\mathcal{L}_i = \partial_\mu R \, \bar{\ell}_i \gamma^\mu \ell_i \,
\sum_{k \, j \, l } \frac{\mbox{Im}\left[
    h^\dagger_{k i} h_{i l} h^\dagger_{k
      j} h_{j l} \right]}{3 M_k M_l} I_{[k l] },
\end{equation}
where $R$ is the Ricci scalar and $I_{ij } = I(M_i,M_j)$ is a loop-factor depending on the sterile masses $M_i$ and $M_j$ in the corresponding diagram and which was computed in full detail in \cite{McDonaldShore:Dec2015}. As described in refs\text{.} \cite{McDonaldShore:Dec2015,McDonald:2014}, this modifies the dispersion relations of leptons and antileptons to
\begin{equation}\label{dispersion}
\left(p_\mu \pm  \partial_\mu R \,
\sum_{k, \, j, \,   l} \frac{\mbox{Im}\left[
    h^\dagger_{k i} h_{i l} h^\dagger_{k
      j} h_{j k} \right]}{3 M_k M_l} I_{[k l] } \right)^2 = 0 ,
\end{equation}
This energy splitting together with $\Delta L = 2$ and $\Delta L =1$ processes drives the system towards a non-zero B-L asymmetry, independently of the departure of sterile neutrinos from equilibrium. For cosmological spacetimes, isotropy and homogeneity mean that spatial derivatives of $R$ vanish and eq. (\ref{dispersion}) leads to an equilibrium B-L to photon ratio of the form 
\begin{equation}\label{NL1}
N^{\text{eq}}_{B-L} = \frac{ \pi^2 \dot{R}}{2 \zeta (3) T} \sum_{i,\, j} \frac{\mbox{Im}\left[ K^2_{i j} \right]}{18 M_i M_j} I_{[ i j]},
\end{equation}
where $K_{ij} = (h^\dagger h)_{ij}$. In this sense, we have a mechanism satisfying all three Sakharov conditions \cite{Sakharov}, the first two of which (particle number and CP violation) are inherited from the usual see saw mechanism. The third - usually stated as a departure from equilibrium - is provided by the time-dependence of the background itself, whose dynamical nature is probed by the lepton screening cloud.

\subsection{Radiation-dominated FRW Cosmology}

In the first part of this Letter, we consider leptogenesis in the conventional radiation-dominated FRW phase of the
evolution of the Universe.  Later, in section 5, we consider earlier times characterised by gravitational sources with more general
equations of state. For radiation dominance, the time variation of the Ricci scalar is
\begin{equation}\label{Riccidot}
\dot R = \sqrt{3} \sigma^{3/2} (1-3w)(1+w)\frac{T^6}{M_p^3}  ,
\end{equation}
where $\sigma = \pi^2/30 g_* $ and $g_* \simeq  106.75$ counts the number of relativistic degrees of freedom in the plasma. Classically, the equation of state parameter $w$ is equal to 1/3 for radiation, and so the expression (\ref{Riccidot}) vanishes. However, trace-anomalies in the gauge sector give $(1-3w) \simeq 10^{-1}$ \cite{Davoudiasl:2004gf}, allowing for $\dot{R} \neq 0$. Combining eqs. (\ref{NL1}) and (\ref{Riccidot})  we arrive at
\begin{align}\label{YLT}
N^{\text{eq}}_{B-L} &\simeq \frac{\sqrt{3}\pi^2 \sigma^{3/2}(1-3w)(1+w)}{36\zeta(3)} \frac{T^5}{M_p^3}
\sum_{i,\, j } \frac{\mbox{Im}\left[ K^2_{i j} \right]}{M_i M_j} I_{[i j]}.
\end{align}

A full description of the general theory of this gravitational leptogenesis mechanism and the calculation
of the equilibrium asymmetry $N_{B-L}^{\rm eq}$ was given in \cite{McDonaldShore:Dec2015}.  In that work, we also made a preliminary 
estimate of the gravitationally induced baryon asymmetry $\eta_B$ based on the assumption that the 
lepton number violating interactions, which maintain the asymmetry at its equilibrium value, freeze out
for temperatures $T_D$ for which $z_D = M_1/T_D \sim 1$. In order to achieve the observed value for $\eta_B$,
we were then led to consider very high sterile neutrino masses and decoupling temperatures at the limits
of existing physical bounds. However, as we demonstrate here, a complete dynamical analysis using the 
full $\Delta L = 2$ reaction cross-section shows that decoupling in fact occurs for significantly smaller
values of $z_D$. Inspection of (6) then makes it clear that the observed asymmetry is achieved for 
lower, conventional values of $M_1 \sim 10^{10} - 10^{11}$ GeV with correspondingly lower decoupling 
temperatures.

Since our interest in ref.~\cite{McDonaldShore:Dec2015} was in the gravitational leptogenesis mechanism itself, we did not discuss the
original mechanism whereby the out-of-equilibrium asymmetric decay rates 
$\Gamma(N \rightarrow \bar{\ell}\bar{\phi}) \neq \Gamma(N\rightarrow \ell\phi)$ of sterile neutrinos 
in the region $z \sim 1$ contribute directly to the B-L asymmetry. Here, we consider the coupled Boltzmann 
equations involving both mechanisms and discuss in some detail the parameter space of the high-energy 
Yukawa phases in which one or other mechanism dominates in determining the final cosmological 
baryon asymmetry.

\section{The Boltzmann Equation}\label{evolution}
We now study the Boltzmann equation to take into account the effect both of sterile neutrino decays and gravitational effects. We shall work in the hierarchical limit where $M_1 \ll M_2 \ll M_3$, so that the dynamics is dominated by the lightest sterile neutrino $N_1$, in which case the relevant Boltzmann equation is (see, \textit{e.g.}, \cite{Pedestrians})
\begin{align}
\frac{dN_{N_1}}{d z} &= - D \left( N_{N_1} - N^{\text{eq}}_{N_1} \right),  \label{Boltzmann1} \\
\frac{dN_{B-L}}{d z} &= - D \varepsilon_1 \left( N_{N_1} - N^{\text{eq}}_{N_1} \right)  - W \left(N_{B-L} - N^{\text{eq}} _{B-L} \right),  \label{Boltzmann2}
\end{align}
where each of the number densities is normalised by the photon density and where $z = M_1/T$. This is the standard set of coupled Boltzmann equations encountered in lepto/baryogenesis (see \textit{e.g.}, \cite{Pedestrians,Buchmuller,Kolb:1979qa}) except that now, due to the gravitational interactions, we have $N^{\text{eq}}_{B-L}\neq 0$ in the RHS of (\ref{Boltzmann2}) in the washout term. Conventionally one has $N^{\text{eq}}_{B-L}=0$ and so any lepton asymmetry generated whilst the sterile neutrinos are in equilibrium is washed out. However, if one takes into account gravitational effects, a lepton asymmetry can be maintained even when $N_{N_1} = N^{\text{eq}}_{N_1}$. 

The CP asymmetry in the decays and inverse decays of sterile neutrinos is characterised by 
\begin{equation}
 \varepsilon_1= \frac{\Gamma(N_1 \rightarrow \ell \phi) -\Gamma(N_1 \rightarrow \overline{\ell} \overline{\phi}) }{\Gamma(N_1 \rightarrow \ell \phi) +\Gamma(N_1 \rightarrow \overline{\ell} \overline{\phi}) } ,
\end{equation}
given in terms of $M_i$ and $h_{ij}$ by \cite{FukYan,Buchmuller}
\begin{equation}\label{decaysYF1}
\epsilon_i  = - \frac{1}{8\pi} \sum_{j \neq i} \frac{\mbox{Im}[ K^2_{i j}]}{K_{ii}}\left[ f\left(\frac{M_j^2}{M_i^2}\right) + g\left(\frac{M_j^2}{M_i^2}\right) \right],
\end{equation}
where
\begin{align}
f(x) &= \sqrt{x}\left( 1  -(1+x) \ln \left(\frac{1+x}{x} \right)\right), \quad
g(x) = \frac{\sqrt{x}}{1-x}.
\end{align}
For a large hierarchy, $x\gg 1$,
\begin{align}
f(x) &\sim  - \frac{1}{2\sqrt{x}}, \quad\quad\quad
g(x) \sim - \frac{1}{\sqrt{x}} .
\end{align}
We shall return to the form of $\varepsilon_1$ in subsequent sections. 

The various reaction rates can be parametrised in terms of the standard quantity $K = \tilde{m}_1/m_*$  \cite{Pedestrians,BuchPlum,BuchBarPlum}  given by
\begin{equation}
\tilde{m}_1 = v^2 \frac{  K_{11} }{M_1}, \qquad m_* = 8 \pi \left( \frac{\pi^2 g_*}{ 90}  \right)^{1/2} \frac{v^2}{M_p} \simeq 10^{-3} ~\text{eV},
\end{equation}
where $\tilde{m}_1$ characterises the strength of the Yukawa interactions and $v = 174$ GeV is the electroweak scale. The quantity $D$ can then be written as
\begin{equation}
D = \frac{\Gamma_{\text{tree}} (N_1 \rightarrow \ell \phi)}{z H} = K z \frac{K_1(z)}{K_2(z)},
\end{equation}
and corresponds to the $N_1 \rightarrow \ell \phi$ tree-level thermal decay width. W is the ``washout term", so-called because when gravitational effects are neglected, $N^{\text{eq}}_{B-L}=0$ and any lepton asymmetry established before the decays of sterile neutrinos is destroyed. The washout term consists of two parts:
\begin{equation}\label{Wtotal}
W = W_{ID} + 2 W_{\Delta L = 2}.
\end{equation}
The first is given by the tree-level inverse decay rate \cite{Pedestrians}
\begin{equation}
W_{ID} =\frac{\Gamma \left( \ell \phi \rightarrow N_1\right)}{z H} =\frac{1}{4} K z^3 K_1(z).
\end{equation}
The second part corresponds to $\Delta L = 2$ binary scatterings $\ell \phi \leftrightarrow  \overline{\ell} \overline{\phi}$ in the $s$- and $u$-channel, and $\ell \ell \leftrightarrow  \phi \phi$ and $\overline{\ell} \, \overline{\ell} \leftrightarrow \overline{\phi}\, \overline{\phi}$ in the $t$-channel. The reaction rates for these processes are given by the quantity $W = \Gamma/zH$, with
\begin{equation}\label{Woriginal}
W = \frac{1}{64 (2\pi)^3} \frac{1}{T^2} \int^\infty_0 ds s^{1/2} K_1 \left(\frac{\sqrt{s}}{T}\right) \frac{1}{s}\left|\mathcal{M}(s) \right|^2,
\end{equation}
where
\begin{equation}
\left|\mathcal{M}(s) \right| = \int^0_{-s} du \left|\mathcal{M} (s,u) \right|^2
\end{equation}
is the $u$-averaged amplitude for the process in question. The amplitudes for $s,u$ and $t$ processes are denoted by the subscripts $+$ and $t$ respectively and take the form
\begin{align}\label{amplitude}
\left| \mathcal{M}_{\Delta L =2} (s)\right|_{+ , \,t}^2& =   2 s^2
\Bigg\lbrace  \frac{K_{11}^2}{M_1^2} F_{+ , \,t}(s)   - 6 \sum_{i \neq 1} \frac{\mbox{Re}\left(K_{1i}^2\right)}{M_1 M_i}  G_{+ , \,t}(s)  \nonumber \\
& + 3 \sum_{j \neq 1} \frac{ \mbox{Re} \left( K_{ij} ^2 \right)}{M_i M_j}
\Bigg\rbrace .
\end{align}
Introducing the variables 
\begin{equation}
c = \frac{K_{11}}{8\pi}, \qquad x = \frac{s}{M_1^2},
\end{equation}
the functions $F$ and $G$ are given by \cite{Pedestrians,BuchPlum}
\begin{align}
F_+ & = \frac{1}{(1-x)^2 + c^2}  - \frac{\pi}{c} \delta(1-x) \nonumber \\
&+ \frac{2}{x} - \frac{2}{x^2} \left( 1 + \frac{x^2 - 1}{(x-1)^2+c^2}\right)  + \frac{2(x-1)}{x \left( (1-x)^2 +c^2 \right)} ,\nonumber \\
\nonumber \\ 
G_+& = \frac{1}{x} + \frac{1}{2}\frac{x-1}{ (1-x)^2 + c^2 } - \frac{x+1}{x^2} \ln (x+1),
\end{align}
and
\begin{align}
F_t & = \frac{2}{x+1} + \frac{2}{x(x+2)} \ln(x+1) ,\nonumber \\
G_t& = -\frac{1}{x} \ln(x+1).
\end{align}
The delta function subtraction in the first line for $F_+$ represents the real intermediate state subtraction from the $s$-channel. This is to avoid the well-known double counting problem \cite{Pedestrians,Kolb:1979qa, BuchPlum} where one  over-counts the number of $N_1 \leftrightarrow \ell \phi$ processes by including them in the $s$-channel $N_1$ exchange. Only with this subtraction does the Boltzmann equation take the correct form, whereby no asymmetry can be generated when $N_{N_1}  = N^{\text{eq}}_{N_1}$. Of course, the whole point of our new mechanism is that $N_{B-L}^{\text{eq}} \neq 0$ and so it \textit{is} possible to generate an asymmetry when the sterile neutrinos are in equilibrium, but in the limit where $N_{B-L}^{\text{eq}} \rightarrow 0$ we should still recover the traditional form of the Boltzmann equation.

Our next task is to parametrise the amplitude (\ref{amplitude}) in terms of neutrino parameters. Firstly we note that
\begin{equation}
 \sum_{i,j = 1,2,3} \frac{ \mbox{Re} \left( K_{ij} ^2 \right)}{M_i M_j} = \frac{\overline{m}^2}{v^4},
\end{equation}
where $\overline{m}^2 = m_1^2 + m_2^2 + m_3^2$ is the sum of the neutrino mass-squares. After a little algebra we can also write
\begin{equation}\label{K1i}
 \sum_{i \neq 1 } \frac{ \mbox{Re} \left( K_{1 i} ^2 \right)}{M_1 M_j}  = \frac{\tilde{m}_1}{v^4 }\left(  \sum_i x_i m_i   - \tilde{m}_1 \right)
\end{equation}
where $x_i$ are $O(1)$ parameters discussed in sec. \ref{CP}. We make the standard choice in the literature \cite{Pedestrians} and set $\mbox{Re}(\tilde{h}_{31}^2) = \mbox{Re}(\tilde{h}_{21}^2) =0$, or equivalently, $x_2=x_3=0$. Equation (\ref{xisum}) then implies $x_1 = m_1/\tilde{m}_1$ and the RHS of (\ref{K1i}) simplifies to  $\left( m_1^2   - \tilde{m}_1^2 \right)/v^4$. Admittedly, this choice is somewhat arbitrary and its main aim is really to reduce the number of free variables, allowing for a simpler parametrisation of the theory. We shall work in this regime for the remainder of this Letter. Putting this together, the amplitudes become
\begin{align}
\left| \mathcal{M}_{\Delta L =2}\right|_{+ , \,t}^2 =  & \frac{2 s^2}{v^4} 
\Bigg[
\tilde{m}_1^2 F_{+ , \,t}(s) + 6 ( G_{+ , \,t}(s) + 1) \left(  m_1^2  - \tilde{m}_1^2 \right) \nonumber \\
&+ 3(\overline{m}^2 - \tilde{m}_1^2)
\Bigg],
\end{align}
allowing us to write eq. (\ref{Woriginal}), after a little manipulation, as  
\begin{align}
&W_{+,\, t} =\frac{ z^3 }{32 \pi^2}  \frac{ m_* M_1}{v^2}  \int_0^\infty d x\, \,  x^{3/2} \, K_1 \left( z \sqrt{x} \right) \nonumber \\
&
 \left[ 
K^2 F_{+ , \,t}(x) + 6 ( G_{+ , \,t}(x) + 1) \left(K^2 - \frac{m_1^2}{m_*^2} \right)+ 3\left(\frac{\overline{m}^2}{m_*^2}- K^2 \right)
\right].
\end{align}
For fixed SM neutrino masses, the amplitude becomes a function of essentially two variables\footnote{Note that $c$ can be written as $c = m_* M_1 K/(8 \pi v^2)$.}  $M_1$ and $K$, which ultimately depend on the details of the high-energy theory. A short calculation also shows that the delta function term in $F_+$ gives a contribution $-W_{ID}$ to $W_{\Delta L = 2}$. 

\begin{figure}
\centering
\includegraphics[scale=0.6]{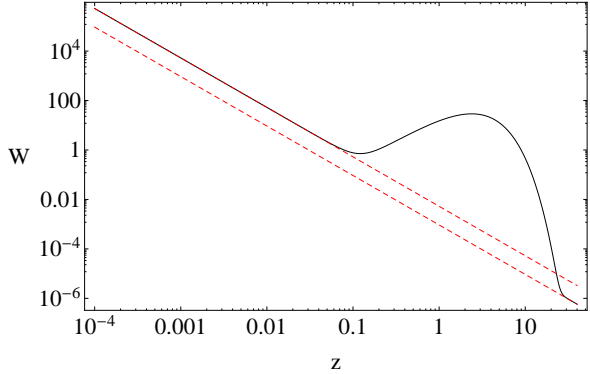}
\caption{Evolution of $W= 2 W_{\Delta L=2} + W_{ID}$ for $K=100$ and $M_1 = 10^{10} \text{GeV}$. The dashed lines show agreement with the asymptotic behaviour for small and large $z$ given by eqs. (\ref{asymp1}),  (\ref{smallzt}) and (\ref{smallz+}).}
\label{Wtot}
\end{figure}

Making the substitution $y = x/z^2$ in the integral, we arrive at
\begin{align}
W_{+,\, t} =&\frac{ 1}{32 \pi^2}  \frac{ m_* M_1}{v^2} \frac{1}{z^2}  \int_0^\infty d y\, \,  y^{3/2} \, K_1 \left( \sqrt{y} \right) \nonumber \\
& \Bigg[ 
K^2 F_{+ , \,t}\left(\frac{y}{z^2} \right) + 6  \left( G_{+ , \,t}\left(\frac{y}{z^2} \right) + 1\right) \left(K^2 - \frac{m_1^2}{m_*^2} \right) \nonumber \\
& + 3\left(\frac{\overline{m}^2}{m_*^2}- K^2 \right)
\Bigg].
\end{align}
Since $F_{+,t}(x), G_{+,t}(x) \rightarrow 0$ as $x\rightarrow \infty$, we see that in the high temperature limit $z \rightarrow 0$, $W_{+,\, t}$ takes the form
\begin{align}\label{asymp1}
&W_{+,\, t}( z \ll 1) \simeq \frac{ 3 }{\pi^2 }  \frac{ m_* M_1}{v^2} \frac{1}{z^2}  
 \left[   \frac{\overline{m}^2}{m_*^2} + K^2 -  \frac{2 m_1^2}{m_*^2} \right],
\end{align}
where we used the result $\int dy y^{3/2}K_1(y) = 32$. 
Similarly, at low temperatures $F_t(0)=3$, $G_t(0)=-1$ leading to
\begin{equation}\label{smallzt}
W_{t}( z \gg 1)  \simeq \frac{ 3 }{\pi^2 }  \frac{ m_* M_1}{v^2} \frac{1}{z^2}  \frac{\overline{m}^2}{m_*^2}.
\end{equation}
Since $F_+(0) = (3 + c^2)/(1 + c^2)$ and  $G_+(0) = -(2+c^2)/(2 (1 + c^2))$, we also have
\begin{align}\label{smallz+}
&W_{+}( z \gg 1) \simeq \frac{ 3 }{\pi^2 }  \frac{ m_* M_1}{v^2} \frac{1}{z^2}  
 \left[  \frac{1 }{v^4} \frac{\overline{m}^2}{m_*^2} +   \frac{1 }{3 m_*^2 v^4} \frac{c^2}{1+ c^2} K^2 \right].
\end{align}
Given that\footnote{The narrow width approximation means that $c = {(h^\dagger h)_{11}}/8\pi = \Gamma_{N_1}/M_1 \ll 1$. This ensures consistency in treating the sterile neutrinos as quasi-stable particle states in the Boltzmann equation.} $c \ll 1$, the second term is sub-dominant, so that to leading order the asymptotic form of eq. (\ref{smallz+}) is the same as (\ref{smallzt}). The contributions to W in eq. (\ref{Wtotal}) are shown in figure \ref{Wtot}, where we took $\overline{m} = \Delta m_{31}^2 + \Delta m_{21}^2 + 3m_1^2 \simeq   \Delta m_{\text{sol}}^2 + \Delta m_{\text{atm}}^2 $, setting $m_1\simeq 0$. 

\section{Parametrising the CP violation}\label{CP}
The fundamental source of CP violation is of course the Yukawa phases contained in $h_{ij}$, or more specifically, the quantities $\mbox{Im}\left(K_{ij}^2\right)$ which control the strength of CP violation both in the lepton propagator and $N^{\text{eq}}_{B-L}$ and also in the decays of sterile neutrinos via $\varepsilon_1$.  One might ask to what extent the CP violation in these two sectors is linked, and also how much each is constrained by low-energy neutrino physics. For hierarchical sterile neutrinos, $M_1 \ll M_2 \ll M_3$ we find that
\begin{equation}\label{decaysYF2}
\varepsilon_1  \simeq  - \frac{1}{8\pi} \sum_{j \neq 1} \frac{\mbox{Im}[ K^2_{1 j}]}{K_{11}} \left( \frac{M_1}{M_j} \right),
\end{equation}
which after a little algebra can be re-written in terms of light neutrino parameters as \cite{BuchBarPlum} 
\begin{equation}\label{decaysYF3}
\varepsilon_1  \simeq  \frac{3}{16\pi} \frac{M_1}{v^2} \sum_{i \neq 1} \frac{\Delta m _{i1}^2}{m_i} \frac{\mbox{Im}\left(\tilde{h}_{i1}^2 \right)}{\left(\tilde{h}_{i1} \right)_{11}}.
\end{equation}
We can parametrise the CP violation in this quantity by using the parameters $z_i$ defined as
\begin{equation}\label{zequation}
\frac{\tilde{h}_{i 1}^2}{(\tilde{h}^\dagger \tilde{h})_{11}} = z _i = x_i + i y_i\, ,
\end{equation}
where  $\sum_{i} \left | z_i \right| = 1$ and $\tilde{h}$ is the mass-eigenstate Yukawa coupling given by $\tilde{h} = U  h$ where $U$ is the PNMS matrix. This satisfies (using the formalism of \cite{Pedestrians} and \cite{Ibarra})
\begin{equation}
\tilde{h}_{ij}  = \frac{1}{\sqrt{m_i M_j}} \Omega_{i j},
\end{equation}
where the see saw formula $\tilde{h}^2_{ij}v^2/M_j = m_i$ implies that $\Omega$ is orthogonal and therefore satisfies $\left(\Omega^T \Omega\right)_{11} =1$. This implies that 
\begin{equation}\label{yconstraint}
\frac{y_1}{m_1} + \frac{y_2}{m_2} + \frac{y_3}{m_3}=0,
\end{equation}
and
\begin{equation}\label{xisum}
\frac{\tilde{m}_1}{m_1} x_1 +\frac{\tilde{m}_1}{m_2} x_2 + \frac{\tilde{m}_1}{m_3} x_3=0.
\end{equation}
Hence the strength of CP violation in $N_1$ decays can be neatly parametrised as
\begin{equation}\label{epsilon1y}
\varepsilon_1 = \frac{3}{16 \pi} \frac{M_1}{v^2}\left( \frac{\Delta m_{21}^2}{m_2} y_2+ \frac{\Delta m_{31}^2}{m_3} y_3\right).
\end{equation} 

One might now ask whether the size of $\varepsilon_1$, or more specifically the quantities $y_i$, uniquely constrain the CP violation appearing in
\begin{equation}\label{CPB-L}
N^{\text{eq}}_{B-L} = \frac{ \pi^2 \dot{R}}{2 \zeta (3) T} \sum_{i \, j} \frac{\mbox{Im}\left[ K^2_{i j} \right]}{18 M_i M_j} I_{[ i j]}.
\end{equation}
The answer to this question is no, as we now explain. Firstly, one should note that ``CP violation" only really makes sense in the context of a particular process, since a given scattering amplitude or decay channel is determined not only by the Yukawa phases in $h_{ij}$, but also by the combinations of masses $M_i$ involved in the relevant diagrams. In this sense, there will be certain regions of parameter space for which CP violation in one process is strong and simultaneously weak in another. For instance, $\varepsilon_1$ depends only on the Yukawa couplings via the quantity $\sum_{j} \mbox{Im}( K_{ij}^2) /M_j$, but this is invariant under the transformation 
\begin{equation}\label{transformation}
\mbox{Im}\left[ K^2_{i j} \right]  \rightarrow  \mbox{Im}\left[ K^2_{i j} \right] +M_* \frac{\epsilon_{i j k }}{M_k},
\end{equation}
where $M_*$ is an arbitrary energy scale. This leaves $\varepsilon_1$ fixed, but changes $\mbox{Im}\left[ K^2_{i j} \right] $ and therefore the size of CP violation in (\ref{CPB-L}), in which $I_{[i j]}$ depends on a completely different combination of masses from those appearing in 
$\varepsilon_1$.

The sterile mass-dependent factor $I_{[ij]}$ was found in ref.~\cite{McDonaldShore:Dec2015} by explicit evaluation 
of the curvature dependence of the two-loop Feynman diagrams in figure 1 to be
\begin{equation}\label{Iasymptotic}
I_{[ij]} \sim  \frac{1}{(4\pi)^4}  \left(\frac{M_j^2}{M_i^2}\right)^p  \ln \left( \frac{M_j^2}{M_i^2} \right),
\end{equation}
in the large hierarchy limit $M_j \gg M_i$. The equilibrium asymmetry is therefore
\begin{equation}\label{NBLeq}
N^{\text{eq}}_{B-L} \simeq \frac{ \pi^2 \dot{R}}{2 \zeta (3) T} \sum_{j > i} \frac{\mbox{Im}\left[ K^2_{i j} \right]}{18 M_i M_j} 
\left( \frac{M_j^2}{M_i^2} \right)^p \ln\left(\frac{M_j^2}{M_i^2}\right) \frac{1}{(4\pi)^4}.
\end{equation}

The dependence on the sterile mass hierarchy is parametrised here by the index $p$. 
In ref.~\cite{McDonaldShore:Dec2015}, strong but not conclusive evidence was found for a hierarchy enhancement with $p=1$.
Analysing the Feynman diagrams in the weak gravitational field limit by attaching gravitons to the sterile neutrino propagators
yields four diagrams, three of which may be explicitly evaluated and give $p=0$.\footnote{Note that in ref.~\cite{McDonaldShore:Dec2015},
``diagram 3'' was incorrectly stated to have $p=1$. However, this leading behaviour in fact cancels leaving a final contribution 
with $p=0$, the same dependence as diagrams 1 and 2. We thank T. Shindou and S. Shirai for bringing this to our attention.}
The fourth is significantly more complex and a complete evaluation has yet to be carried through. However, it was shown in
\cite{McDonaldShore:Dec2015} that $p=1$ contributions (but no higher) arise throughout and barring a final cancellation
will provide the dominant behaviour. In the following section, where we consider a conventional radiation-dominated FRW cosmology, 
we therefore assume a hierarchy enhancement with $p=1$.  In section 5 we compute the gravitationally-induced
lepton asymmetry in an alternative cosmological background with the more conservative choice $p=0$ to show that the 
observed asymmetry may still be obtained even without a power-law hierarchy enhancement.

Returning to (\ref{NBLeq}), we therefore see that constraining the size of $\varepsilon_1$ still leaves the three quantities $\mbox{Im}\left[ K^2_{13} \right],\mbox{Im}\left[ K^2_{23}\right]$ and $\mbox{Im}\left[K^2_{12}\right]$ undetermined, so that  the size of $N^{\text{eq}}_{B-L}$ is not fully constrained in terms of $y_i$ of eq.~(\ref{zequation}) . In this sense, the gravitational effect is sensitive to different details of the high-energy see-saw physics compared to the usual delayed decay picture and is less constrained by SM neutrinos. Of course, in future work it could be interesting to see what other low-energy observables could be used to constrain the combination of masses appearing in eq. (\ref{NBLeq}).

\section{Evolution of the lepton asymmetry}

We now describe the solution of the Boltzmann equations (\ref{Boltzmann1}) and  (\ref{Boltzmann2}),  highlighting the different
leptogenesis scenarios that occur depending on the relative strength CP-violation from gravity and decays, which can be dialed independently by virtue of the transformation  (\ref{transformation}).  For our present purposes, we assume that the $\mbox{Im}[K_{ij}^2]$ are of roughly the same order of magnitude and that they realise a fixed value of $\varepsilon_1$. Therefore, assuming $M_1 \ll M_2 \ll M_3$, the sum in eq. (\ref{NBLeq}) is dominated by the $N_1,N_3$ contribution, giving
\begin{equation}
N^{\text{eq}}_{B-L} \simeq \frac{ \pi^2 \dot{R}}{36\zeta (3) T (4\pi)^4}  \frac{\mbox{Im}\left[ K^2_{1 3} \right]}{ M_1 M_3} 
\left( \frac{M_3^2}{M_1^2} \right)^p \ln\left(\frac{M_3^2}{M_1^2}\right).
\end{equation}

\begin{figure} [t!]
\includegraphics[scale=0.58,trim={2cm 1.5cm 0 2cm}]{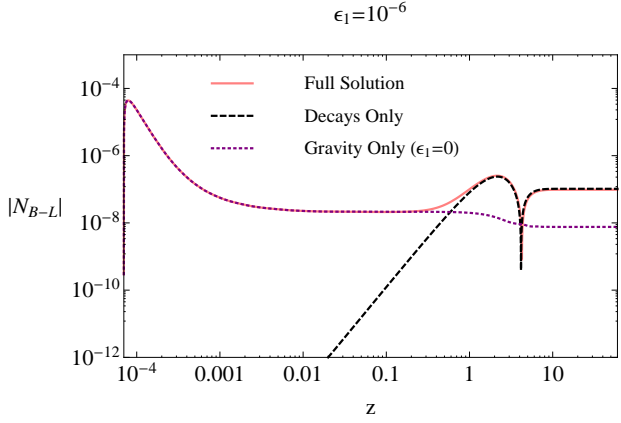}
\caption{Plot of the evolution of $|N_{B-L}|$ with $z$ for $K=1$, $M_1 = 10^{10}\text{GeV}$ with $\varepsilon_1 = 10^{-6}$ and $\mbox{Im}(K_{13}^2)  /(4\pi)^2= 10^{-6}$, $M_3 = 10^{16}$,
assuming a hierarchy enhancement with $p=1$. In the full solution (pink), we see that at early times, there is a gravitationally induced asymmetry, but the $\varepsilon_1 D(N_1 - N_1^{\text{eq}})$ term dominates in the Boltzmann equation as we approach $z=1$ and the asymmetry is determined solely by CP violating decays, with no memory of the gravitational effects at early times. The purple dotted curve, which includes only gravitational effects and neglects decays by setting $\varepsilon_1=0$, shows that decays have no effect until $z \sim 1$.}
\label{epsilon1}
\end{figure}

\begin{figure}[t!]
\includegraphics[scale=0.61,trim={2cm 1.5cm 0 0}]{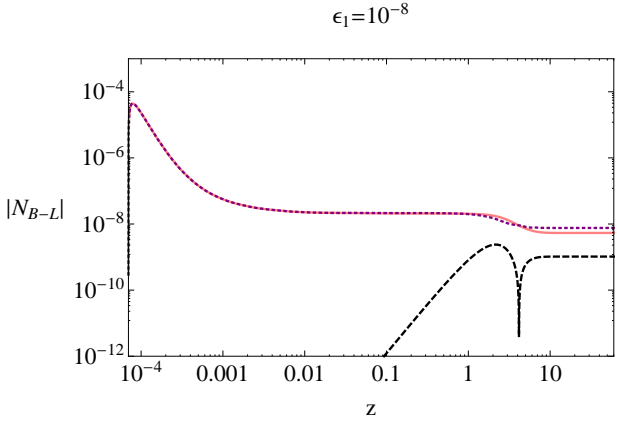}
\caption{The other parameters are the same as figure \ref{epsilon1}, but we now take $\varepsilon_1 = 10^{-8}$. For this value of $\varepsilon_1$, the full solution is solely dominated by gravitational effects (pink curve), i.e. the decays have no effect on the relic asymmetry. This can be clearly seen by comparison with the dotted purple curve, which neglects decays entirely by setting $\varepsilon_1=0$, and shows that the full solution is essentially independent of decays. From the black dashed curve, we see that taking into account decays alone does not give an accurate representation of the true solution.   }
\label{epsilon1other}
\end{figure}

We now examine what happens when both decays and gravitational effects are present (figures \ref{epsilon1} and \ref{epsilon1other}), by considering different values of $\varepsilon_1$, whilst keeping CP-violation in the gravitational sector fixed.
Of course, it should be noted that our ability to dial the two effects independently is due to the sterile mass-dependence unique to the curved-space two-loop diagrams in figure \ref{2loop}. Ultimately the contribution to dispersion relations can be traced to the \textit{real} part of these \textit{curved-space} self-energies. In contrast, the combination of masses appearing in $\varepsilon_1$, is a result of the \textit{imaginary} parts of \textit{flat-space} self-energies, which come from the relevant cuts through two-loop diagrams and relate to decay rates. The analysis of \cite{McDonaldShore:Dec2015} was crucial to understand the parametric details of the gravitational mechanism and the important asymptotic behaviour $I_{[ij]} \sim \left(M_j^2/M_i^2\right)^p \ln(M_j^2/M_i^2)$, which contrasts with that of $\varepsilon_1$. It is this richness of parameter space which leads to the distinct leptogenesis scenarios described below.

In all cases, even if we start from a vanishing initial net lepton number at high temperatures, the system very rapidly attains its
gravitationally-induced equilibrium asymmetry $N_{B-L}^{\text{eq}}(z) \neq 0$. The asymmetry then tracks this equilibrium value as the 
Universe cools.
As the corresponding rate for the lepton number-violating interactions falls (see figures \ref{Wtot} and \ref{WK1}), the system can no longer 
follow the extremely rapid $1/z^5$ decrease in $N_{B-L}^{\text{eq}}$ (see eq.(\ref{YLT})) and the asymmetry freezes out. 
The region of $z$ at which this decoupling takes place depends on the sterile neutrino mass $M_1$ and $K$, which control the washout 
coefficient $W$. In the scenarios illustrated here, decoupling takes place 
for small values of $z$, significantly below the scale $z \sim 1-10$ at which the effects of the $N_1$ resonance in $W$ and the 
$N_1$ decays are felt.
In the first scenario (figure \ref{epsilon1}), we consider maximal $\varepsilon_1 \simeq 10^{-6}$ 
(setting $y_2 \simeq0,~y_3\simeq 1$ in  (\ref{epsilon1y})) as in the standard delayed-decay picture. 
Then with the parameters shown, including the hierarchy enhancement $p=1$, 
since the asymmetry generated by the out-of-equilibrium $N_1$ decays is larger than the gravitational effect 
and occurs later (for $z \gtrsim 1$), the gravitationally-induced asymmetry is taken over by decays, and the system evolves according to the 
conventional decay scenario with no memory of the early-time gravitational effects.

A scenario where $\varepsilon_1$ is smaller is shown in figure \ref{epsilon1other}. In this case, although the sterile neutrino decays
do generate an asymmetry as usual, this effect is smaller than the gravitationally-induced asymmetry after freeze-out. Remarkably, 
therefore, in this scenario the final asymmetry is completely determined by the gravitational mechanism, with the decays playing no
significant role. This alters our understanding of the parameter space of leptogenesis, showing that regions which were previously believed 
to give an asymmetry in terms of decays are actually dominated by the gravitational mechanism.

\subsection{Gravity only: the extremal case $\varepsilon_1=0$}
Since our main interest here is in illustrating the mechanism of gravitational leptogenesis, we now study in detail the extremal case where the 
CP-violating decay parameter $\left |\varepsilon_1 \right| \simeq 0$ is minimal. In this case, only the washout scatterings contribute and the 
Boltzmann equation for $N_{B-L}$ simplifies radically:
\begin{equation}\label{simpleBoltz}
\frac{dN_{B-L}}{d z} =  - W \left(N_{B-L} - N^{\text{eq}}_{B-L} \right). 
\end{equation}
Note here that in the region of interest, $z\ll 1$, a good approximation to the washout term for neutrino parameters 
$m_1 \ll \bar{m}$ is given from (\ref{asymp1}) by 
\begin{equation}\label{approxW}
W = \frac{\alpha}{z^2},  ~~~~~~~~~~~ \alpha = \frac{6}{\pi^2} \frac{M_1 m_*}{v^2} \left(\frac{\bar{m}^2}{m_*^2} + K^2\right).
\end{equation}

As we now see, this scenario is readily realised by choosing opposite signs for the Yukawa phases in (\ref{decaysYF2}), (\ref{epsilon1y}). This places a constraint on the high energy physics of the form 
\begin{equation}\label{epsilonzero}
 \varepsilon_1 \simeq 0 \implies M_3\mbox{Im}\left[ K^2_{ 1 2} \right] + M_2\mbox{Im}\left[ K^2_{1 3} \right] \simeq 0,
\end{equation}
or equivalently, from eq. (\ref{epsilon1y}),
\begin{equation}
 \frac{\Delta m_{21}^2}{m_2} y_2+ \frac{\Delta m_{31}^2}{m_3} y_3 \simeq 0.
\end{equation}
Even with this restriction, there still remains much freedom in the choice of CP violation in the quantities $\mbox{Im}[K_{ij}^2]$ contained in (\ref{YLT}) - for instance, eq. (\ref{epsilonzero}) places no constraints on the phases of $K_{23}^2$. For simplicity, we set $\mbox{Im}[K_{23}^2]=0$,
then from eqs.~(\ref{Iasymptotic}) and (\ref{epsilonzero}) we find
\begin{align}
& \sum_{i \, j } \frac{\mbox{Im}\left[ K^2_{ij} \right]}{M_i M_j}  I_{[i,j]} \nonumber \\
&\simeq  \frac{1}{M_1 M_3} \frac{\mbox{Im} \left[K_{13}^2 \right]}{(4\pi)^4}   \left(\frac{M_3^2}{M_1^2} \right)^p  
\left\{ \ln \left( \frac{M_3^2}{M_1^2}\right) - \left(\frac{M_2^2}{M_3^2}\right)^p \ln \left( \frac{M_2^2}{M_1^2}\right)\right\}.
\end{align}
so that if $M_1 \ll M_2 \ll M_3$ we have
\begin{equation}
\sum_{i,j } \frac{\mbox{Im}\left[ K^2_{ij} \right]}{M_i M_j}  I_{[i,j]} \simeq  
\frac{ \mbox{Im} \left[K_{13}^2 \right] }{(4\pi)^4} \frac{1}{M_1 M_3}  \left(\frac{M_3^2}{M_1^2} \right)^p \ln \left( \frac{M_3^2}{M_1^2}\right).
\end{equation}
Notice that the size of the CP asymmetry is enhanced by the hierarchy between $M_3$ and $M_1$. In what follows, we treat $\mbox{Im}[K_{13}^2]$ as a free parameter (subject to the constraint (\ref{epsilonzero})) controlling the strength of CP violation. Putting this together 
and taking $p=1$, we find
\begin{align}\label{equilibriumvalue}
N^{\text{eq}}_{B-L} &\simeq \frac{\sqrt{3}\pi^2 \sigma^{3/2}(1-3w)(1+w)}{36\zeta(3)} \nonumber \\
&\times   \frac{1}{z^5} \left(\frac{M_1}{M_p}\right)^3   \frac{M_3}{M_1} \ln \left(\frac{M_3^2}{M_1^2} \right) \frac{\mbox{Im}[K_{13}^2]}{(4\pi)^4} \nonumber \\
&\equiv \frac{\beta}{z^5}.
\end{align}
\begin{figure}
\includegraphics[scale=0.53]{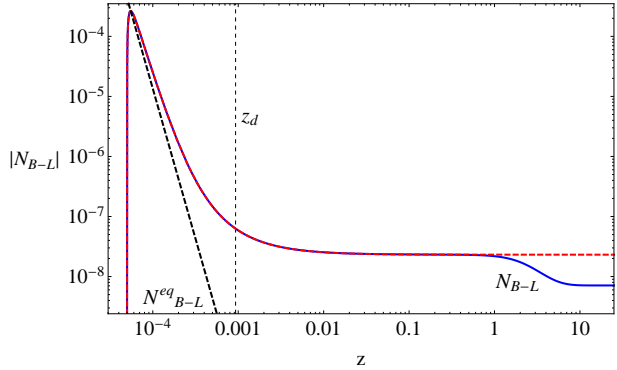}
\caption{Solutions to the Boltzmann equation (\ref{simpleBoltz}), for $K=1$ for fixed $M_1=10^{10} \text{GeV}$ and  $\mbox{Im}(K_{13}^2) /(4\pi)^2 = 10^{-6}$, $M_3 = 10^{16}\text{GeV}$, taking $p=1$. The blue line shows the numerical solution, the red the analytic solution, valid at early times whilst the black dashed line gives the equilibrium curve. The vertical dashed line shows the value $z_d$ where $\Gamma/H\simeq 1$.}
\label{boltzsol}
\end{figure}

\begin{figure}
\includegraphics[scale=0.64,trim={0cm 0 0 0.6cm}]{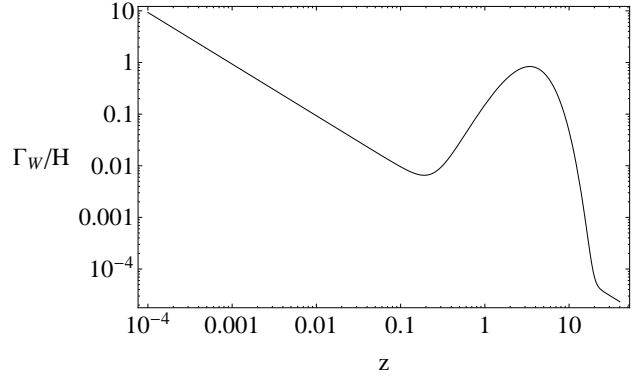}
\caption{The ratio $\Gamma_W/H$  with $K=1$ and $M_1 = 10^{10}$ GeV.}
\label{WK1}
\end{figure}

\begin{figure}[h]
\includegraphics[scale=0.65]{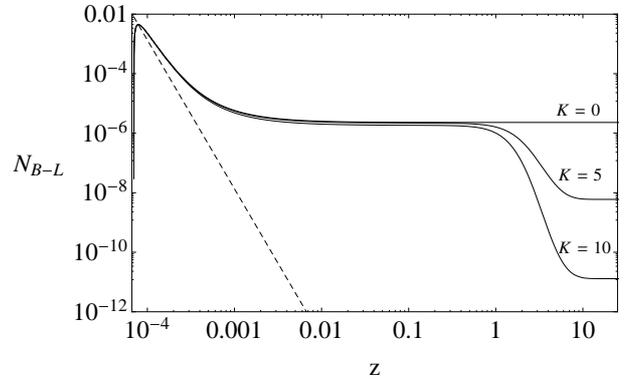}
\caption{Some of the solutions corresponding to figure \ref{relic} with $\mbox{Im}[K_{13}^2]/(4\pi)^2=10^{-4}$.}
\label{Ksolutions}
\end{figure}

The corresponding solution of the Boltzmann equation (\ref{simpleBoltz}) in this scenario is shown in figure \ref{boltzsol}.
In this case, following the freeze-out of the asymmetry from its equilibrium value\footnote{A little more insight into these numerical solutions
follows from solving the Boltzmann equation (\ref{simpleBoltz}) in the $z\ll 1$ region analytically. From (\ref{approxW}) and
(\ref{equilibriumvalue}), we have 
\begin{equation*}\label{boltzapprox}
N_{B-L}'(z) =  \frac{\alpha}{z^2}\left( N_{B-L} (z) - \frac{\beta}{z^5} \right), \qquad z \ll 1,
\end{equation*}
which admits an analytic solution with asymptotic value 
\begin{equation*}
N_{B-L}^f \simeq 120 \frac{\beta}{\alpha^5}.
\end{equation*}
This solution is plotted alonside the full numerical solution in figure \ref{boltzsol} and is a useful guide in scanning
the parameter space of $M_1$ and $K$.}, the only further new feature is the late-time reduction of
$N_{B-L}$ in the region $z \sim 1-10$ which is due to the contribution to $W$ near the 
$N_1$ resonance. This raises the value of $W$ and pulls the asymmetry back, albeit only slightly with the parameter choice in 
figure \ref{boltzsol}, in the direction of the equilibrium value. This is also apparent from figure \ref{WK1}, where it is clear that 
$\Gamma_W/H$ once again becomes of order 1 in this region. The size of this late-time reduction in $N_{B-L}$ depends on the choice
of parameters $M_1$ and $K$, in particular increasing sharply with $K$ as illustrated in figure \ref{Ksolutions}.

\begin{figure}
\includegraphics[scale=0.48,trim={0.8cm 0 0 0}]{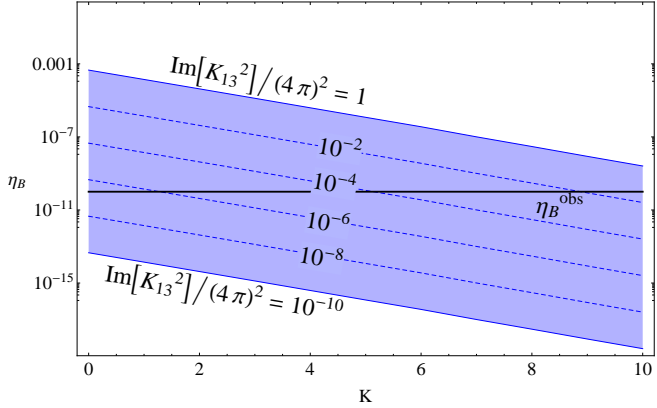}
\caption{Numerical results for the final baryon-to-photon ratio as a function of $K$ for $M_1 = 10^{10}$GeV and $M_3 = 10^{16}$GeV,
assuming a hierarchy enhancement with $p=1$.  The amount of CP
violation is varied by taking $\mbox{Im}(K_{13}^2) /(4\pi)^2 =1 -  10^{-10}$ (series of dashed lines), which simply shifts the overall scaling of $N_{B-L}$, as can be seen from eqs. (\ref{simpleBoltz}) and (\ref{equilibriumvalue}).}
\label{relic}
\end{figure}

The key observation, however, is that even in this model with the CP-violating parameters chosen such that the sterile neutrino decays
produce a negligible asymmetry, the gravitational leptogenesis mechanism on its own can produce the observed cosmological baryon 
asymmetry for an otherwise conventional choice of see-saw neutrino parameters. For example, in figure \ref{boltzsol} the sterile neutrino
masses were chosen to be $M_1 = 10^{10}$ GeV, $M_3 = 10^{16}$ GeV and $K=1$, with $\mbox{Im}(K_{13}^2) /(4\pi)^2 = 10^{-6}$.
The corresponding value for the final relic baryon asymmetry is given by 
\begin{equation}
\eta_B = \frac{1}{f} C_{\text{sph}} N^\text{f}_{B-L},
\end{equation}
where $f = 2387/86$ is a photon production factor and $C_{\text{sph}} = 28/70$ is the sphaleron efficiency factor \cite{Pedestrians,Buchmuller}.
Clearly, the observed asymmetry, $\eta_B \simeq 10^{-10}$ can be obtained for a significant range of the parameters $M_1$, $M_3$, 
$\mbox{Im}(K_{13}^2) $ and $K$. In figure \ref{relic}, we illustrate the dependence of $\eta_B$ on 
$\mbox{Im}(K_{13}^2) /(4\pi)^2$ and $K$ for fixed $M_1$, $M_3$.

\section{Alternative Cosmological Backgrounds}

Finally, we relax the choice of a conventional radiation-dominated FRW background and, following \cite{Davoudiasl:2004gf},
consider a more general scenario in which we allow the gravitational background to be sourced by matter characterised by
an equation of state with arbitrary parameter $w$. Specifically, we consider an isotropic, homogeneous geometry whose matter source
has an energy density $\rho \sim a^{-3(w+1)}$, where $a$ is the scale factor of the Universe. Potential sources, for example
scalar fields, giving rise to different values of $w$ are discussed further in ref.~\cite{Davoudiasl:2004gf}.
The plasma in which leptogenesis takes place corresponds in this scenario to a {\it sub-dominant} radiation component 
for which $\rho_R \sim a^{-4}$ with temperature $T$ satisfying $\rho_R = \sigma T^4$. The onset of radiation dominance occurs at a critical temperature $T_*$ where $\rho \simeq \rho_R$,
and leptogenesis takes place in the pre-radiation dominance phase of the evolution above $T_*$.

We can then parametrise both matter and radiation energy densities in terms of the plasma temperature $T$ and critical temperature $T_*$
as follows:
\begin{equation}\label{new1}
\rho = \sigma T_*^4 \left(\frac{T}{T_*}\right)^{3(1+w)} , ~~~~~~~~ \rho_R = \sigma T^4 \ .
\end{equation}
The curvature for $T > T_*$ is sourced by $\rho$, so that here
\begin{equation}\label{new2}
\dot{R} = \sqrt{3} (1-3w) (1+w) \frac{\rho^{3/2}}{M_p^3} \ ,
\end{equation}
which may be written as
\begin{equation}\label{new3}
\dot{R} = \sqrt{3} (1-3w) (1+w) \sigma^{3/2} \frac{M_1^6}{\gamma^6 M_p^3} \left(\frac{\gamma}{z}\right)^{9(1+w)/2} \ ,
\end{equation}
where we have introduced the parameter $\gamma = M_1/T_*$.
This gives rise to an equilibrium lepton-to-photon ratio
\begin{align}\label{new4}
N^{\text{eq}}_{B-L} \simeq &\frac{\sqrt{3}\pi^2 \sigma^{3/2}(1-3w)(1+w)}{36\zeta(3)} \nonumber \\
 & ~~~~ \times~\frac{M_1^5}{\gamma^5 M_p^3} \left(\frac{\gamma}{z}\right)^{9(1+w)/2 -1} ~
\sum_{i,\, j } \frac{\mbox{Im}\left[ K^2_{i j} \right]}{M_i M_j} I_{[i j]} \ ,
\end{align}
which may be compared with (\ref{YLT}).

\begin{figure} 
\includegraphics[scale=0.61,trim={2cm 1.5cm 0 2cm}]{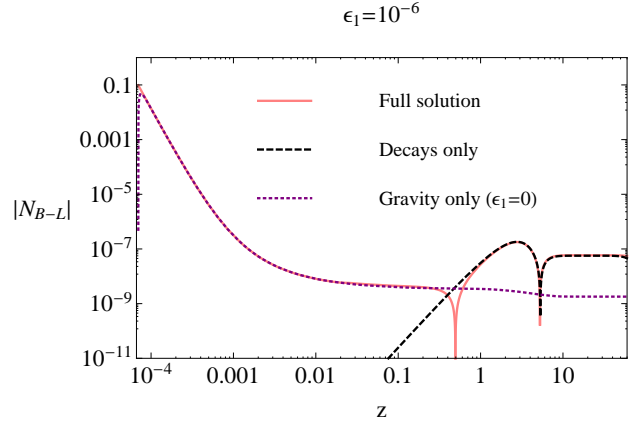}
\caption{Plot of the evolution of $|N_{B-L}|$ with $z$ for $K=1$, $M_1 = 5 \times 10^{11}\text{GeV}$ with $\varepsilon_1 = 10^{-6}$ and $ \mbox{Im}(K_{13}^2)   /(4\pi)^2= 10^{-4}$, $M_3 = 10^{16}$ in the generalised cosmological model. The cosmological parameters chosen were $\gamma = M_1/T_R = 40$ and $w=0.5$
and the hierarchy parameter was taken here as $p=0$. The essential features are the same as illustrated in figure \ref{epsilon1} for the radiation-dominated spacetime.}
\label{epsilon1NEW}
\end{figure}

The analysis of the Boltzmann equations now goes through essentially as before, showing all the same qualitative features.
The main quantitative difference arises from the temperature dependence of the equilibrium asymmetry, which from
(\ref{new4}) falls off as $z^{-9(1+w)/2 +1}$, depending on the parameter $w$ characterising the source of the gravitational background.
Moreover, unlike the radiation-dominated scenario, where for a standard sterile neutrino sector we required the hierarchy
enhancement $p=1$ in order to reproduce the observed baryon asymmetry, in this model the freedom to choose the
parameters $w$ and $\gamma$ means that it is possible to obtain $\eta_B \simeq 10^{-10}$ even without this enhancement.
To illustrate this, we take $p=0$ in the plots shown in this section.

\begin{figure}
\includegraphics[scale=0.62,trim={0.72cm 0.2cm 0 2cm}]{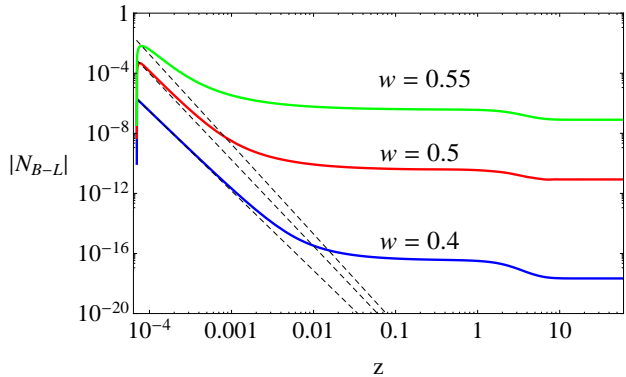}
\caption{Solutions to the Boltzmann equation (\ref{simpleBoltz}) for the generalised cosmological model, 
with $K=3$, $M_1=5 \times 10^{11} \text{GeV}$,
$\mbox{Im}(K_{13}^2) /(4\pi)^2 = 10^{-4}$, $M_3 = 10^{16}\text{GeV}$ and taking $p=0$. 
Results are shown for $\gamma = 40$ and three illustrative values of $w$. 
The corresponding equilibrium curves are drawn as dashed lines. }
\label{boltzsolNEW}
\end{figure}

The evolution of the asymmetry $|N_{B-L}|$ is shown in figures
\ref{epsilon1NEW} and \ref{boltzsolNEW}. In figure \ref{epsilon1NEW},
the analogue of figure \ref{epsilon1}, we illustrate the competition between the gravitational and decay mechanisms
for leptogenesis with the CP violating parameter $\varepsilon_1 = 10^{-6}$ for a cosmological model with $w = 0.5$ and
$\gamma = 40$, where decoupling from $N^{\text{eq}}_{B-L}$ takes place for temperatures with $z\lesssim 10^{-3}$. 
Clearly for smaller values of the CP violating parameter, the situation again resembles 
figure \ref{epsilon1other} with the decays being irrelevant and the final asymmetry dominated by the gravitational mechanism.
Figure \ref{boltzsolNEW} shows the dependence of the decoupling temperature and the final asymmetry on the equation of state parameter $w$. 
Unsurprisingly, $|N_{B-L}|$ is seen to be extremely sensitive to $w$, reflecting the power dependence in 
$N^{\text{eq}}_{B-L} \sim z^{-9(1+w)/2 +1}$.

Overall then, we see that in this cosmological scenario in which leptogenesis occurs before the onset of radiation dominance,
where the background spacetime is sourced by matter with an as yet undetermined value of $w$, the observed baryon asymmetry 
may still be obtained for a significant range of neutrino and cosmological parameters even in the absence of a hierarchy
enhancement of the Feynman diagram factor $I_{[ij]}$ characterising the gravitationally-induced lepton number asymmetry.

\section{Conclusions}
In this Letter, we have presented a detailed study of the dynamics of lepton number generation in the early Universe, taking into 
account both the conventional out-of-equilibrium decays of the sterile neutrinos in the see-saw model and our new mechanism
of gravitational leptogenesis \cite{ RIGLletter,McDonaldShore:Dec2015}. This has demonstrated clearly for the first time
that this gravitational mechanism is indeed capable of generating the observed baryon asymmetry $\eta_B \simeq 10^{-10}$.

This study, which sheds new light on traditional perspectives in leptogenesis, involved a full numerical analysis of the coupled
Boltzmann equations, modified to include the non-vanishing equilibrium asymmetry generated at two-loop order by the gravitational
interactions. The parameter space of high-energy Yukawa phases was explored fully, showing that the CP violation in the gravitational 
and sterile neutrino decay sectors can be dialled independently. Whether the final asymmetry is determined by the gravitational or
decay effects is then controlled by the size of the CP-violating decay parameter $\varepsilon_1$. 
In particular, even in the limit of minimal $\varepsilon_1 \simeq 0$, we showed that the observed value of $\eta_B$ may be obtained for 
otherwise standard choices of neutrino parameters in the see-saw model. This establishes radiatively-induced gravitational leptogenesis
as a viable mechanism for explaining the matter-antimatter asymmetry of the Universe.

\section*{Acknowledgements}
JIM would like to thank D. De Boni for useful conversations about kinetic theory. We are grateful to the STFC who funded this research under grants ST/K502376/1 and ST/L000369/1.
\\

\end{document}